\documentclass[preprint,prmaterials,aps,amsmath,amsfonts]{revtex4-2}

\usepackage[english]{babel}
\makeatletter
\def\selectlanguage#1{}
\makeatother
\usepackage{graphicx}
\usepackage{amsmath,amssymb}
\usepackage{colortbl}
\usepackage{subcaption}
\usepackage{ulem}
\usepackage{xcolor}          
\usepackage{booktabs}  
\usepackage{setspace}  

\begin{document}

\begin{abstract}
Calculating diffusion rates of point defects in materials typically relies on the harmonic approximation to estimate migration free energies. However, anharmonic effects can have a large impact on diffusion properties, and explicitly accounting for them is usually computationally demanding and difficult to achieve in practice.
In this work, we investigate the role of anharmonic effects on defect migration in UO$_2$ and PuO$_2$ using the potential of average force integration (PAFI) method. Fully anharmonic migration free energies are computed for several cation and anion defect types, using the Cooper-Rushton-Grimes (CRG) potential and a recently developed machine learning spectral neighbour analysis potential (SNAP) for UO$_2$. Results are systematically compared to harmonic estimates based on attempt frequencies and the Debye approximation.
Our results reveal that the validity of the harmonic approximation strongly depends on the defect type and the underlying potential, with significant deviations observed in several cases. In particular, defect migration barriers are found to decrease strongly with increasing temperature (up to 1 eV between 0 and 1\,200 K), and anharmonic contributions can substantially modify migration entropies and, consequently, diffusion coefficients. Comparing defect migration in UO$_2$ and PuO$_2$ using the CRG potential reveals that PuO$_2$ has lower migration enthalpies at 0~K for all considered defects, but this is compensated by higher attempt frequencies, resulting in similar overall jump frequencies in UO$_2$ and PuO$_2$. These findings provide insight into the limitations of commonly used approximations and highlight the importance of anharmonic effects for predictive modeling of diffusion in nuclear fuels as well as in other classes of materials.
\end{abstract}

\title{Fully anharmonic calculations of the free energy of migration of point defects in UO$_2$ and PuO$_2$}

\author{D. G. Frost}
\affiliation{CEA, DES, IRESNE, DEC, Cadarache, Saint-Paul-Lez-Durance, F-13108, France}

\author{J. Bouchet}
\affiliation{CEA, DES, IRESNE, DEC, Cadarache, Saint-Paul-Lez-Durance, F-13108, France}

\author{M. C. Marinica}
\affiliation{Université Paris-Saclay, CEA, Service de Recherche en Corrosion et Comportement des Matériaux, SRMP, Gif-sur-Yvette, F-91191, France}

\author{C. Lapointe}
\affiliation{Université Paris-Saclay, CEA, Service de Recherche en Corrosion et Comportement des Matériaux, SRMP, Gif-sur-Yvette, F-91191, France}

\author{J. B. Maillet}
\affiliation{CEA, DAM, DIF, F‑91297 Arpajon, France}
\affiliation{Université Paris‑Saclay, CEA, Laboratoire Matière en Conditions Extrêmes, 91680 Bruyères‑le‑Châtel, France}

\author{L. Messina}
\affiliation{CEA, DES, IRESNE, DEC, Cadarache, Saint-Paul-Lez-Durance, F-13108, France}

\date{\today}   
\maketitle
	
\pagebreak
\section{Introduction}

Diffusion is a fundamental transport process that controls the redistribution of atoms, defects, and charge carriers in solids. It plays a central role in determining the structural evolution, stability, and functional properties of materials, influencing phenomena from phase transformations to electronic performance. A quantitative understanding of diffusion is therefore essential for both fundamental studies and the design of advanced materials.

UO$_2$ is the most common fuel material used in nuclear reactors today and is likely to remain so for the foreseeable future. Plutonium content in UO$_2$ progressively increases with fuel irradiation, and mixed (U,Pu)O$_2$ fuels are fabricated to reuse fissile Pu recovered from spent nuclear fuel. Determining the atomic diffusion is important throughout the lifecycle of nuclear fuel, from manufacturing to waste storage and disposal. During manufacture the diffusion of oxygen governs the temperature and time used for sintering which can vastly alter the costs of production of fuel~\cite{lay_role_1969,fuhrman_lowtemperature_1963,frost_controlling_2024}.
During reactor operations and subsequent storage, self-diffusion and diffusion of defects and fission products can alter the thermophysical and mechanical properties of the fuel~\cite{andersson_simulations_2012,fisher_microstructure_2002,rest_fission_2019,forsberg_diffusion_1985}. Of particular note in the case of high burnup fuels, the diffusion of fission gases can cause changes in internal pressure of the fuel rods, potentially leading to fuel ruptures~\cite{fisher_microstructure_2002,rest_fission_2019}. Diffusion in UO$_2$ and PuO$_2$ is essential for developing physics-based predictive models and simulations of nuclear fuel behaviour. 

Comprehensive experimental studies on diffusion processes in UO$_2$ and PuO$_2$ were largely carried out by Matzke, along with extensive reviews of the literature~\cite{matzke_formation_1997,matzke_atomic_1990,matzke_radiation_1983,matzke_rim_1992}.
Matzke's work built upon their own early work as well as the work on UO$_2$ and PuO$_2$ by Bayoglu \textit{et al.}~\cite{bayoglu_mesure_1983}, Marin and Contamin~\cite{marin_uranium_1969}, Kim \textit{et al.}~\cite{kim_oxygen_1981} and Belle~\cite{belle_oxygen_1969}.
These works determined experimentally the diffusion coefficients, migration barriers and activation energies for self-diffusion, point defects, fission gas diffusion coefficients in UO$_2$ and PuO$_2$ and interdiffusion in (U,Pu)O$_2$. These studies serve as the benchmark against which newer research using density functional theory (DFT) and experiments is typically evaluated. More recent studies, focusing on oxygen diffusion in UO$_2$ and PuO$_2$ have been undertaken by Kato \textit{et al.}~\cite{kato_oxygen_2023,watanabe_oxygen_2023}.

Whilst the diffusion coefficients of various defects in UO$_2$ and PuO$_2$ have been found experimentally from the above studies, there are still unanswered questions regarding diffusion mechanisms and underlying physical phenomena~\cite{dorado_first-principles_2012}. There have been many studies that have examined self diffusion in UO$_2$ using first principles. The migration barriers and formation energies of point defects in UO$_2$ have been calculated in notable works by Petit \textit{et al.}~\cite{petit_point_1998}, Crocombette \textit{et al.}~\cite{crocombette_plane-wave_2001} and Dorado \textit{et al.}~\cite{dorado_atomistic_2010,dorado_first-principles_2011,dorado_advances_2013,wang_activation_2013}. Crocombette \textit{et al.}~\cite{crocombette_plane-wave_2001} and Freyss \textit{et al.}~\cite{freyss_point_2005} found similar formation energies calculated using GGA+$U$ and LDA+$U$ functionals, respectively, which compare favourably with experimental results in a comparison done by Liu \textit{et al.}~\cite{liu_first-principles_2012}. Dorado \textit{et al.} found good agreement between calculations of the migration barriers of oxygen vacancies and interstitials, however, the agreement was poor for uranium vacancies/interstitials~\cite{dorado_first-principles_2012,dorado_first-principles_2011,dorado_advances_2013}. Further first principles studies have looked at the diffusion of fission products, fission gases and other actinides in UO$_2$~\cite{brillant_study_2008,brillant_investigation_2009,gupta_ab_2009,gryaznov_ab_2009,gryaznov_helium_2010,perriot_diffusion_2015}. A comprehensive review of DFT studies in UO$_2$ has been compiled by Liu \textit{et al.}~\cite{liu_first-principles_2012}. Further research in this area has been conducted by Maillard \textit{et al.}~\cite{MAILLARD2022} and Neilson \textit{et al.}~\cite{Neilson2024}  who investigated the effects of stoichiometry. PuO$_2$ has also been studied, however, not nearly as comprehensively as for UO$_2$ with most computational studies occurring within the last decade~\cite{freyss_ab_2006,hernandez_dftu_2016,nakamura_first-principles_2018,singh_first_2024,tang_first_2023, Neilson2021}.

First principles calculations can capture complex interactions including charge transfer, however, the high computational costs make it prohibitive to explore large phase spaces. Therefore, there have been numerous studies utilising empirical potentials to determine diffusion mechanisms and coefficients. Considerable work in the diffusion domain of actinide oxides has been completed using the Cooper-Rushton-Grimes (CRG)~\cite{cooper_many-body_2014} empirical potential owing to its predictive capabilities and wide range of available fission products and alternative actinides~\cite{cooper_many-body_2014,cooper_vacancy_2014,cooper_modeling_2015,cooper_defect_2018,cooper_thermophysical_2015,perriot_atomistic_2019,owen_diffusion_2023,liu_small_2021}. Whilst other potentials have been used in the past, studies have found the CRG potential to provide better agreement with DFT+$U$ for formation and migration energies compared to earlier empirical potentials, specifically Liu \textit{et al.} compared values from DFT for small interstitial clusters in UO\(_2\) with results from the Morelon~\cite{morelon2003new} and CRG potentials~\cite{liu_small_2021,kuksin_calculation_2014,govers_comparison_2007,arima_molecular_2010}.

Analysis of the available literature has shown that there is very little information regarding the migration barriers found in PuO\(_2\) one of which looked at the diffusion of oxygen point defects~\cite{zhao2025modeling}. All the currently available studies for UO\(_2\) and PuO\(_2\) use DFT or empirical potentials to determine the migration energy at 0~K. Those studies are often combined with phenomenological models for diffusion at higher temperatures, typically by approximating the attempt frequency as the average atomic vibration frequency (Debye frequency). Calculating the attempt frequencies directly for each defect migration rather than using the Debye frequency can improve estimates of the diffusion coefficients at temperatures above 0~K. Given that nuclear fuel operates at temperatures that can exceed 1\,200~K~\cite{fink2000thermophysical}, direct calculation of the migration free energy along with attempt frequencies is valuable information for fuel performance codes.

Two interatomic potentials were chosen for this work, the CRG embedded atom method (EAM)~\cite{cooper_many-body_2014,cooper_development_2016} potential and a machine learning spectral neighbour analysis potential (SNAP) for UO$_2$ recently developed by Dubois \textit{et al.}~\cite{dubois_atomistic_2024}. The CRG potential was selected as it is the most widely used interatomic potential for actinide oxides. However, the CRG potential was not fitted to any diffusion parameters. In contrast, the SNAP potential was chosen for the fact that it was fitted to defect formation energies obtained from DFT, which may result in more accurate migration energy calculations. Nevertheless, neither the CRG nor the SNAP potential were explicitly trained on migration trajectories. While DFT would be the ideal approach for such calculations, its computational cost is a few orders of magnitude higher~\cite{dubois_atomistic_2024, zuo2020performance}. Anionic and cationic vacancies, interstitials and bound Schottky defects (BSDs) in UO$_2$ and PuO$_2$ are studied. These point defects were chosen as they have been frequently studied in the literature and can be compared with DFT and/or experiments and are also the most common under thermal equilibrium and irradiation conditions~\cite{fink2000thermophysical,freyss_point_2005,crocombette2011charge,dorado_advances_2013}. 

In this work, we build on previous studies that have evaluated 0~K migration barriers and explored migration behaviour in UO$_2$ by calculating attempt frequencies and migration free energies. We extend this approach by employing a new machine learning potential, alongside the CRG potential, to perform nudged elastic band (NEB) calculations for key defects. Migration free energies are computed directly using a novel method developed by Swinburne \textit{et al.}~\cite{swinburne_unsupervised_2018} between 0~K to 1\,200~K, incorporating enthalpy and entropy contributions. We compute attempt frequencies directly, enabling comparison between the Debye harmonic approximation and the attempt frequency harmonic approximation of the migration free energy. Jump frequencies are then derived and compared using the three different methods. We also compare the behavior of UO$_2$ and PuO$_2$ using the CRG potential, and benchmark our results against available DFT data and experimental observations. These jump frequencies are directly relevant to modelling irradiated nuclear fuel, where defect concentrations deviate from thermal equilibrium, and can be integrated into fuel performance codes to improve predictions beyond current harmonic approximations.

\section{Methodology}
\label{sec:methodology}	

\subsection{Defect migration free energy and jump frequency}
In this work we wish to calculate the jump frequency of each defect as a function of temperature which can be done by using transition state theory~\cite{mehrer_diffusion_2007}. Within transition state theory the diffusion coefficient, $D$, can be calculated using Eq.~\ref{eqn:1} which shows the relationship between the jump frequency ($\omega$), the concentration of defect $d$ ($C_d$) as well as a correlation factor $f$, a geometric factor based on the crystal structure $g$, and the lattice parameter $a_0$: 

\begin{equation}
\label{eqn:1}
    {D = g f a_0^2\omega C_d} \; .
\end{equation}

In nuclear reactors, the concentration of defects is dominated by irradiation rather than thermal equilibrium. Therefore, we can ignore the defect concentration and geometric factors, focusing solely on the jump frequency. The jump frequency can be calculated using transition state theory expressed as the transition rate in the Eyring–Evans–Polanyi formulation \cite{eyring_activated_1935, evans_applications_1935, Kramers}, in the limit where the transmission coefficient is set to one. This is described by Vineyard~\cite{vineyard_frequency_1957}, the equations derived from which are shown in Eq.~\ref{eqn:2} that shows the relationship between the jump frequency, the attempt frequency, \text{$\nu_0$} and the Gibbs free energy of migration, $G_\mathrm{M}$.  $k_\mathrm{B}$ and $T$ are the Boltzmann constant and temperature, respectively:

\begin{equation}
\label{eqn:2}
    {\omega = \nu_0 \exp \left(- \frac{G_\mathrm{M}}{k_\mathrm{B} T}\right)} \; .
\end{equation}

The normal approach from here is to use the harmonic approximation using the relation ${G_\mathrm{M} = H_\mathrm{M} - TS_\mathrm{M}}$, where $H_\mathrm{M}$ and $S_\mathrm{M}$ are the enthalpy and entropy of migration, respectively. The jump frequency can then be expressed using these relationships in Eq.~\ref{eqn:3}~\cite{mehrer_diffusion_2007}.
\begin{equation}
\label{eqn:3}
    {\omega = \nu_0 \exp\left(\frac{S_\mathrm{M}}{k_\mathrm{B}} \right) \exp\left(-\frac{H_\mathrm{M}}{k_\mathrm{B} T} \right)} \; .
\end{equation}

The harmonic approximation can be used in order to compute the entropic contribution using ${S_\mathrm{M} = k_\mathrm{B} \ln \nu_0}$. The attempt frequency, $\nu_0$, is often not calculated explicitly and instead the Debye frequency is used as an estimate, denoted here as $\nu_\mathrm{D}$. This would then give the following expression for the jump frequency, ${\omega = \nu_\mathrm{D} \exp\left(\frac{S_\mathrm{M}}{k_\mathrm{B}} \right) \exp\left(-\frac{H_\mathrm{M}}{k_\mathrm{B} T} \right)}$ and ${S_\mathrm{M} = k_\mathrm{B} \ln \nu_\mathrm{D}}$. This is the baseline harmonic approximation used in this work, where only $H_\mathrm{M}$ is calculated.

The second approximation uses the attempt frequency in place of the Debye frequency, calculated for each defect using Eq.~\ref{eqn:4} developed by Vineyard~\cite{vineyard_frequency_1957}. This is identical to the Debye harmonic approximation approach but uses the calculated attempt frequency in place of the Debye frequency:

\begin{equation}
\label{eqn:4}
{\nu^0 = \frac{\prod_{i=1}^{3N-3}\nu_\mathrm{IS}}{\prod_{i=1}^{3N-4}\nu_\mathrm{TS}}} \; .
\end{equation}

Equation~\ref{eqn:4} takes the product of all real eigenfrequencies, i.e., those greater than 0 Hz, at the initial state, $\nu_\mathrm{IS}$ and divides them by those at the transition state, $\nu_\mathrm{TS}$. $N$ is the number of atoms in the supercell.

The quantity G$_\mathrm{M}$ found in Eq.~\ref{eqn:2} can be computed directly for moderately complex transitions. The Projected Average Force Integrator (PAFI) method~\cite{swinburne_unsupervised_2018} offers an optimal solution for the fully anharmonic determination of the migration free energy which includes phonon-phonon interactions and temperature dependent effects. PAFI performs constrained sampling on zero-temperature energy-pathway hyperplanes, yielding an analytical estimate of the free energy gradient. This allows accurate determination of free energy barriers, even when the minimum energy path differs from the true minimum free energy path. The PAFI method aims to reconstruct the free energy profile along a reaction path connecting two states via one or more saddle points. PAFI requires a continuous, one-dimensional reaction coordinate, typically initialised using a zero-temperature minimum energy pathway found using NEB calculations~\cite{henkelman_climbing_2000}.

We therefore aim to compare two distinct harmonic approximations, one using the calculated attempt frequency and the other using the Debye frequency as a substitute, against the direct calculations using PAFI. The migration enthalpy contribution $H_\mathrm{M}$ is computed as the energy barrier at 0~K evaluated along the minimum energy path of the transition. The NEB climbing image method described by Henkelman~\cite{henkelman_climbing_2000} was used for these calculations, which additionally provides configurations corresponding to the transition state and the initial state and are used to determine the attempt frequencies. The quantities for $\nu_0$, $H_\mathrm{M}$, and $S_\mathrm{M}$ then allow the Gibbs free energy of migration and the jump frequency to be calculated for the two types of harmonic approximation and compared to those calculated directly using PAFI.

\subsection{Calculation details} 

The LAMMPS package was used for all simulations in this work~\cite{thompson_lammps_2022}. We used the CRG interatomic potential~\cite{cooper_many-body_2014} for calculations in UO$_2$ and PuO$_2$, and a machine-learning SNAP potential developed by Dubois et. al.~\cite{dubois_atomistic_2024} for UO$_2$ only.
The SNAP interatomic potential~\cite{dubois_atomistic_2024} was developed using active training on a dataset derived from DFT+$U$ using a $U$ value of 4.5~eV and a $J$ parameter of 0.54~eV.
The training dataset included different polymorphs, defect formation and migration energies and generalised stacking fault energy surfaces~\cite{dubois_atomistic_2024}. The CRG potential was fit to experimentally available data, in particular the thermal expansion and elastic constants at 0~K~\cite{cooper_many-body_2014}. The particle-particle particle-mesh (PPPM) solver~\cite{hockney2021computer} is employed to compute the long range Coulombic interactions included in both potentials. The short range interactions for the CRG and SNAP potentials are, respectively, cut off at 11.0~\AA~and 5.8~\AA~\cite{cooper_many-body_2014,dubois_atomistic_2024}. Supercells of the given $M$O$_2$ (where $M$ is the cation U or Pu) of 1\,500 atoms in a $5 \times 5 \times 5$ cubic fluorite unit cell arrangement were created and relaxed at 0~K to provide the initial configurations. 

\begin{figure*}[htbp]
	\includegraphics[width=\textwidth, keepaspectratio]{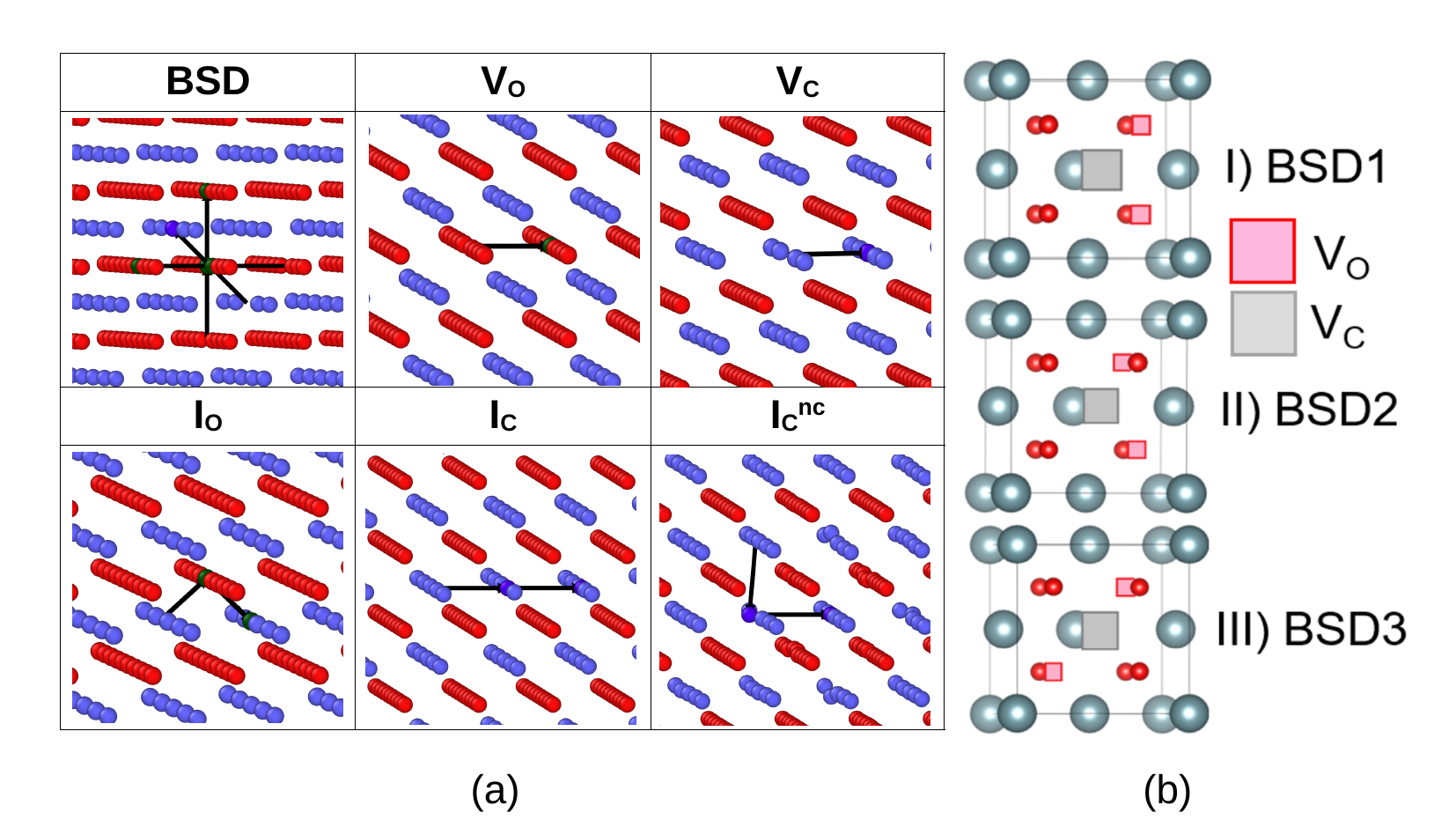}
	\caption{Defect configurations and migration pathways investigated in this work. (a) Migration mechanisms for vacancy and interstitial defects in the fluorite lattice. Diffusing atoms are highlighted in green for oxygen and purple for cations. From left to right: (top) bound Schottky defect (BSD), oxygen vacancy (V$_\mathrm{O}$), cation vacancy (V$_C$); (bottom) oxygen interstitial (I$_\mathrm{O}$), cation interstitial with collinear (I$_C$) and noncollinear (I$_C^\mathrm{nc}$) migration. (b) Bound Schottky defect configurations, with cations shown in grey and oxygen atoms in red, with vacancy positions marked by square symbols.}
	\label{fig:defects}
\end{figure*}

Defects were then added to the relaxed supercells for both the cations and oxygen interstitial/vacancy. Defects constituted a Frenkel pair to maintain overall charge neutrality and stoichiometry on the whole simulation domain. The distance between the interstitial and the vacancy is maximised in the simulation cell so as to avoid interactions between them. The defect which is undergoing a migration will be specified using the following convention: I$_\mathrm{C}$ for a cation interstitial, of which two distinct migration pathways are under investigation, indirect collinear and indirect noncollinear, where the noncollinear pathway is differentiated as I$_\mathrm{C}^\mathrm{nc}$. Both mechanisms are kick-out type mechanisms as direct interstitial migration pathways have been shown to have a migration barrier almost double that of kick-out mechanisms (7.9 eV vs. 4.1 eV~\cite{dorado_first-principles_2012}). Figure~\ref{fig:defects}a shows the migration paths of each defect studied, clearly indicating the difference between the collinear and indirect pathways. The indirect collinear path 
pushes the defect in a single direction while the noncollinear pushes the defect at a right angle to the original path~\cite{mehrer_diffusion_2007}. The cation vacancies will be shown as V$_\mathrm{C}$, oxygen interstitials and vacancies as I$_\mathrm{O}$ and V$_\mathrm{O}$, respectively. Only indirect mechanisms were chosen for the oxygen interstitial pathways due to having migration barriers significantly lower than those associated with direct mechanisms~\cite{dorado_first-principles_2012,dorado_first-principles_2011}. Bound Schottky defects will be shown as BSD. There are three possible BSD configurations depending on the position of the oxygen atoms~\cite{vathonne2014dft+,burr_importance_2017,bathellier_effect_2022}; however, here only the $\langle111\rangle$ BSD (sometimes referred to as BSD3) is calculated as it has the lowest formation and migration energies of the three types of BSD~\cite{burr_importance_2017}. These different configurations are shown in Fig.~\ref{fig:defects}b. To differentiate between potentials and species CRG-UO\(_2\) will refer to UO$_2$ using the CRG potential, CRG-PuO\(_2\) will refer to PuO$_2$ calculations using the CRG potential and SNAP-UO\(_2\) will refer to UO$_2$ using the SNAP potential.

It is important to note that for all calculations there is only a single defect in the supercells, either a Frenkel pair or a BSD. The supercells including defects then underwent an energy minimisation to find the minimum energy configuration at zero pressure for each defect and potential. The resulting configurations correspond to the initial configurations for the NEB calculations. The final configurations were determined by moving the defect to the nearest appropriate site. The minimum energy pathways for each defect were determined using the method outlined by Henkelman \textit{et al.} on climbing image NEB calculations~\cite{henkelman_climbing_2000}, using the implementation found in LAMMPS~\cite{thompson_lammps_2022}. The NEB calculation separates the initial and final configuration and then interpolates a linear path between these two points. Each of these replicas is connected by inter-replica nudging forces which maintain spacing between the images whilst the interatomic forces are minimized to converge on a minimum energy path~\cite{henkelman_climbing_2000}. The convergence threshold for inter-replica forces was set to 0.001~eV$\cdot$\AA$^{-1}$, the number of images used was 19 and the inter-replica nudging force, or KSpring, was set to 1.0 eV$\cdot$\AA$^{-1}$. 

From the NEB calculations, the initial configuration as well as the one corresponding to the saddle point, or transition state, are considered for subsequent calculation of the attempt frequencies. The Phonopy package~\cite{togo_first_2015} was used to generate supercells with finite atomic displacements (0.03~\AA~used here) and generate a force constant matrix. This generated on the order of 10$^3$ supercells for which atomic forces are evaluated using LAMMPS in order to build the force constant matrix. From this force constant matrix the dynamical matrix was evaluated and the density of states and eigenfrequencies were calculated. 
The attempt frequency is then deduced from these results as defined earlier.

The minimum energy pathways generated from the NEB calculations were recalculated using equal energy spacing as described by the LAMMPS neb fix~\cite{thompson_lammps_2022} with 47 images, this is intended to improve the spline fit of the initial pathway. These images were employed to compute the migration free energy as a function of temperature with PAFI~\cite{swinburne_unsupervised_2018}. PAFI performs constrained sampling on hyperplanes of the minimum energy pathway by means of the adaptive biasing force~\cite{comer_adaptive_2015} method. This method continuously adds a biasing force to the equations of motion, allowing for sampling to occur uniformly on a flat free-energy surface, then providing a reliable estimate of the free energy~\cite{comer_adaptive_2015, swinburne_unsupervised_2018}. Unlike harmonic approaches, PAFI captures temperature-dependent anharmonic effects and phonon-phonon interactions, providing a physically meaningful free energy barrier at elevated temperatures. PAFI calculations were performed for each potential and defect at temperatures of 300, 600, 900, and 1\,200~K. The number of sample steps and thermalisation steps were fixed to 1\,000.
The friction parameter was set to 0.01 and the number of samples per hyperplane was set to 1\,920. The thermal expansion coefficients should be supplied to PAFI in order to accurately resize the simulation box before the thermalisation steps. These were determined for each potential and are reported in Table~\ref{table:alpha}. Thermal expansion on the lattice parameter $a(T)$ is computed as a second order expansion in temperature:
\begin{equation}
    a(T) = a_0 \left(1 + \alpha T + \beta T^2 \right) \; .
    \label{eqn:thermal_expansion}
\end{equation}
These were calculated from 0 to 1\,300~K using 50~K temperature intervals and then fit to Eq.~\ref{eqn:thermal_expansion}.

\begin{table}
\caption{Lattice parameters and fitted thermal expansion coefficients (Eq. \ref{eqn:thermal_expansion}) used to account for lattice expansion in PAFI calculations.}
\label{table:alpha}
\setlength{\tabcolsep}{7pt}
\centering
\begin{tabular}{cccc}
    \toprule
	Potential & $a_0$ (\AA) & $\alpha$ ($\times 10^{-6} \mathrm{K}^{-1}$) & $\beta$ ($ \times 10^{-9} \mathrm{K}^{-1}$)\\
	\midrule
	CRG-UO\(_2\) & 5.454 & 8.90 & 1.38 \\
	SNAP-UO\(_2\) & 5.536 & 17.4 & -1.10  \\
	CRG-PuO\(_2\) & 5.381 & 8.58 & 1.57 \\
	\bottomrule
\end{tabular}
\end{table}

\section{Results}
\label{sec:results}
\begin{figure*}[htbp]
 \centering
 \includegraphics[width=\textwidth, keepaspectratio]{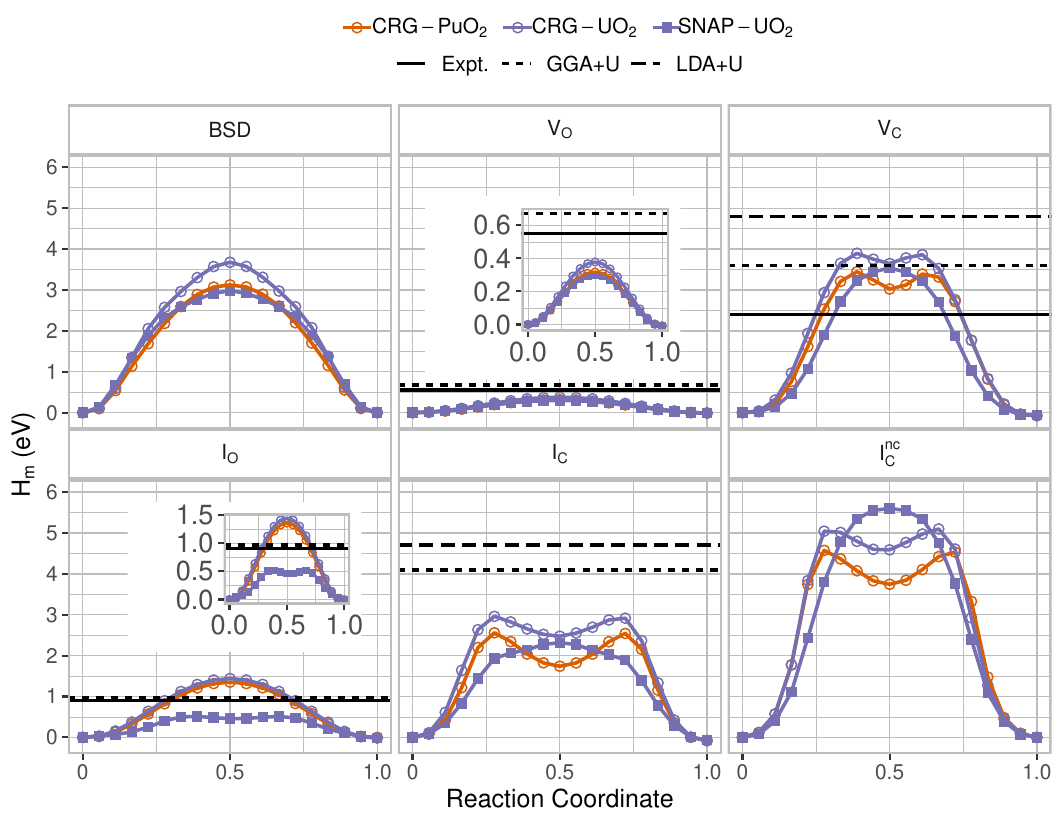}
 \caption{Minimum energy migration pathways obtained using nudged elastic band (NEB) calculations for the CRG-PuO$_2$, CRG-UO$_2$, and SNAP-UO$_2$ interatomic potentials. Migration enthalpy barriers for UO$_2$ reported from DFT calculations (GGA+$U$ and LDA+$U$)~\cite{dorado_first-principles_2011,dorado_first-principles_2012} and experiments~\cite{matzke_atomic_1990} are included for comparison.}
 \label{fig:NEB}
\end{figure*}

 Figure~\ref{fig:NEB} contains the results of the NEB calculations compared with results from literature, where available. These previous results show only the migration enthalpy values, $H_\mathrm{M}$, and are differentiated using different horizontal line types. There are two sets of results from DFT, those found using the generalised gradient approximation (GGA+$U$) and the local density approximation (LDA+$U$). In general, GGA+$U$ provides better accuracy at higher computation cost. The experimental values for the migration enthalpy were taken from Matzke~\cite{matzke_atomic_1990}. Matzke~\cite{matzke_atomic_1990} determined the diffusion coefficients in UO$_2$ at temperatures between 1\,690 and 1\,990~K under different oxygen partial pressures, determining the migration enthalpy of I$_\mathrm{O}$ to be between 0.8 and 1.0 eV, V$_\mathrm{O}$ to be between 0.6 to 0.8 eV and the migration enthalpy of V$_C$ to be approximately 2.4 eV. The values shown in Fig.~\ref{fig:NEB} are the average of those published by Matzke~\cite{matzke_atomic_1990}. The results from Matzke~\cite{matzke_atomic_1990} show a lower migration enthalpy than those calculated using DFT and GGA+$U$ finds lower migration enthalpy than those found with LDA+$U$. The NEB calculations show the change in the enthalpy of migration as the defect moves through the potential energy surface towards the next minima along the reaction pathway. The enthalpy of migration varies widely depending on the type of defect, with the highest enthalpy of migration attributed to the indirect cation migration (I$_C^\mathrm{nc}$) and the lowest for the oxygen vacancy (V$_\mathrm{O}$). The values found using DFT are higher than those found in experiments. DFT results for the cation vacancy, V$_C$, using GGA+$U$ are in excellent agreement with the value found using the CRG potential, however, this doesn't extend to other defects.
 
For all cationic point defects I$_C$, I$_C^\mathrm{nc}$ and V$_C$ using the CRG-UO$_2$ or CRG-PuO$_2$ potential there is a local minima between two saddle points of equal energy. This local minimum is not predicted by the SNAP-UO$_2$ potential. Further investigation of the $I_C^\mathrm{nc}$ defect shows two oxygen atoms being dragged by the moving cation interstitial before breaking the interatomic bonds and reaching a transition (or saddle point) state and then returning to their original positions. The local minima occurs when a new Frenkel pair is formed before the created vacancy is filled by the existing interstitial. The SNAP-UO$_2$ potential doesn't show the same interaction with the local oxygen atoms. The oxygen point defects exhibit migration barriers lower than for point defects with cations. Notably, in all cases the defect migration barrier is lower in PuO$_2$ than in UO$_2$ when comparing between the CRG potentials. Comparing the SNAP-UO$_2$ with the CRG-UO$_2$ potential, the former always predicts lower migration barriers except in the case of the I$_C^\mathrm{nc}$ defect. 

\begin{figure*}[htbp]
\centering
\includegraphics[width=\textwidth, keepaspectratio]{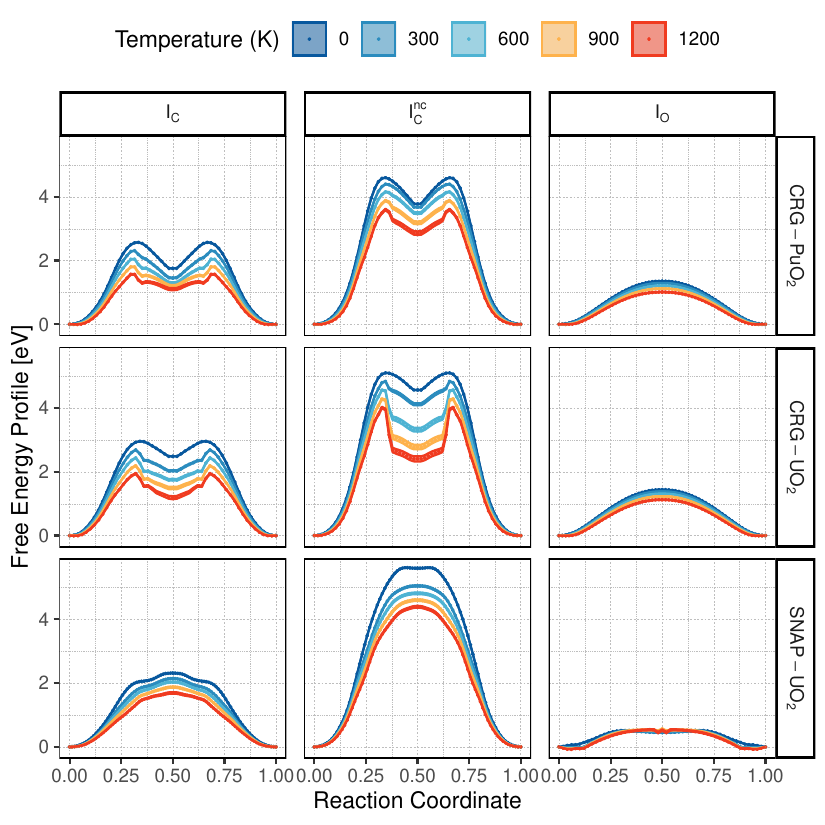}
\caption{Temperature-dependent migration free energy profiles for interstitial defects (cation interstitial I$_C$, cation interstitial with noncollinear mechanism I$_C^\mathrm{nc}$, and oxygen interstitial I$_\mathrm{O}$) computed using the projected average force integrator (PAFI) method between 0 and 1\,200 K with the three interatomic potentials.}
\label{fig:PAFI_ints}
\end{figure*}

Using the initial pathway derived from NEB calculations the Gibbs energy of migration ($\Delta G_\mathrm{M}$) can be determined as a function of temperature using PAFI. This allows for direct determination of $\Delta G_\mathrm{M}$ beyond the use of the harmonic approximation for migration barriers at finite temperatures. Figure~\ref{fig:PAFI_ints} shows the free energy of migration and migration pathway every 300~K from 0~K up to 1\,200~K for the interstitial type defects ($I_C$, $I_C^\mathrm{nc}$ and $I_\mathrm{O}$). The corresponding migration barriers as functions of temperature are reported in Table \ref{table:PAFI-results}. The 0~K pathways from PAFI are in excellent agreement with those determined using the NEB method which verifies that the spline fit of the NEB pathway is well defined. The migration pathways for all defects and potentials maintain much of the features of the original pathway, including in the cases of defects with two saddle points. 

\begin{figure*}[htbp]
    \centering
    \includegraphics[width=\textwidth, keepaspectratio]{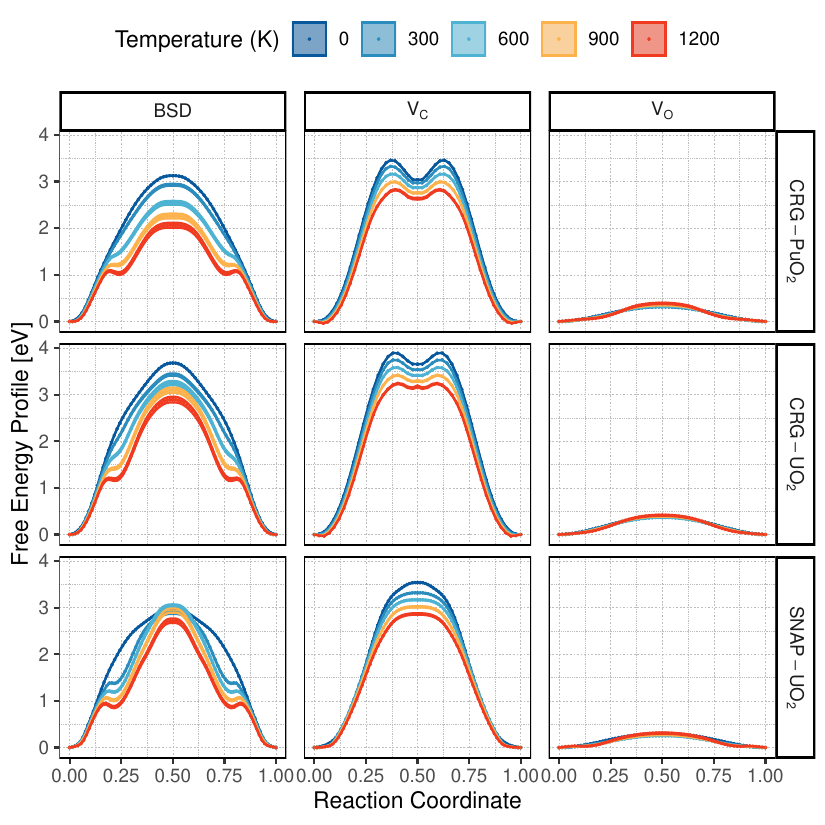}
    \caption{Temperature-dependent migration free energy profiles for vacancy defects (bound Schottky defect BSD, cation vacancy V$_C$, and oxygen vacancy V$_\mathrm{O}$) computed using the projected average force integrator (PAFI) method between 0 and 1\,200 K with the three interatomic potentials.}
    \label{fig:PAFI_vacs}
\end{figure*}

Figure~\ref{fig:PAFI_vacs} contains the migration free energy for each of the vacancy type defects (BSD, $V_C$ and $V_\mathrm{O}$), with the corresponding migration barriers reported in Table \ref{table:PAFI-results}.
The cationic defects, I$_C$, I$_C^\mathrm{nc}$ and V$_C$ tend to maintain a similar local minima in the energy whilst the saddle point energy decreases, resulting in a flattening of the energy pathway. These migration free energies can now be compared with the harmonic approximation, this requires determination of \(S_\mathrm{M} = \mathrm{k_\mathrm{B}} \ln\nu_0\) using the Debye and attempt frequency for $\nu_0$.

\begin{figure}[htbp]
	\centering
    \includegraphics[width=\columnwidth, keepaspectratio]{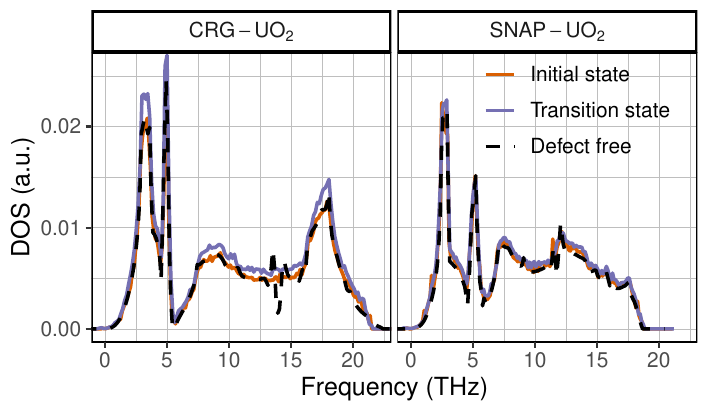}
	\caption{Phonon density of states calculated using the CRG-UO$_2$ and the SNAP-UO$_2$ interatomic potentials. Spectra are shown for the perfect lattice, the initial defect configuration, and the defect transition state (saddle point).}
	\label{fig:DOS_BSD}
\end{figure}

For each pathway the initial configuration and transition configuration (saddle point) have been extracted to determine the associated attempt frequency. 
This was done using the frozen phonon method outlined in the Methodology. The phonon density of states (DOS) was computed and analysed visually. An example for a bound Schottky defect is shown in Fig.~\ref{fig:DOS_BSD} using both the CRG-UO\(_2\) and SNAP-UO$_2$ potentials. Additionally, the DOS for the perfect cell is shown. Differences between the DOS of the perfect and defective lattices are most pronounced at energies between 12-13~THz where the DOS of the defective structure is smoother than the DOS of the perfect lattice. This is the case for both the CRG-UO\(_2\) and SNAP-UO$_2$ potentials. Imaginary frequencies (those below 0 THz) at the transition state are difficult to observe in Fig.~\ref{fig:DOS_BSD}, but they are present upon closer examination of the frequency data. The attempt frequency was calculated from the phonon frequencies, the results of which are found in Table~\ref{table:attfreq} along with the Debye frequencies for each of the defect free species. The bound Schottky defect has the highest attempt frequency of all tested defects, with an attempt frequency greater than 10~THz for all tested potentials. The next highest attempt frequencies are found for the I$_C$ defect, except in the case of the SNAP-UO$_2$ potential. Vacancy defects have the lowest attempt frequencies of the tested defects. There are two defects, namely the cationic point defects I$_C$ and V$_C$, which have very low attempt frequencies when calculated using the SNAP-UO$_2$ potential compared to those from the CRG-UO\(_2\) and CRG-PuO\(_2\) potentials.

\begin{table}
\caption{Attempt frequencies $\nu_0$ (THz) and migration enthalpies $H_\mathrm{M}$ (eV) at 0 K obtained from NEB calculations for each defect and interatomic potential. Debye frequencies $\nu_\mathrm{D}$ (THz) used in the harmonic approximation are also listed.}
\label{table:attfreq}
\centering

\setlength{\tabcolsep}{6pt}

\begin{tabular*}{\columnwidth}{@{\extracolsep{\fill}}c cc cc cc}
\toprule

& \multicolumn{2}{c}{CRG-PuO$_2$}
& \multicolumn{2}{c}{CRG-UO$_2$}
& \multicolumn{2}{c}{SNAP-UO$_2$} \\

\cmidrule(lr){2-3}
\cmidrule(lr){4-5}
\cmidrule(lr){6-7}

Defect 
& $\nu_0$ & $H_\mathrm{M}$
& $\nu_0$ & $H_\mathrm{M}$
& $\nu_0$ & $H_\mathrm{M}$ \\

\midrule

I$_\mathrm{O}$ 
& 1.93 & 1.35 
& 1.63 & 1.44 
& 8.59 & 0.51 \\

I$_C$ 
& 10.0 & 2.56 
& 7.96 & 2.96 
& 0.01 & 2.32 \\

I$_C^{\mathrm{nc}}$ 
& 1.69 & 4.57 
& 0.590 & 5.10 
& 1.32 & 5.60 \\

BSD 
& 60.9 & 3.12 
& 10.85 & 3.68 
& 13.23 & 2.98 \\

V$_\mathrm{O}$ 
& 1.59 & 0.32 
& 1.62 & 0.38 
& 2.38 & 0.30 \\

V$_C$ 
& 1.95 & 3.45 
& 1.70 & 3.90 
& 0.06 & 3.54 \\

\cmidrule(lr){2-3}
\cmidrule(lr){4-5}
\cmidrule(lr){6-7}

$\nu_\mathrm{D}$
& \multicolumn{2}{c}{8.83} 
& \multicolumn{2}{c}{8.02} 
& \multicolumn{2}{c}{8.02}  \\

\bottomrule
\end{tabular*}
\end{table}

\begin{figure*}[htbp]
	\centering
	\includegraphics[width=\textwidth, keepaspectratio]{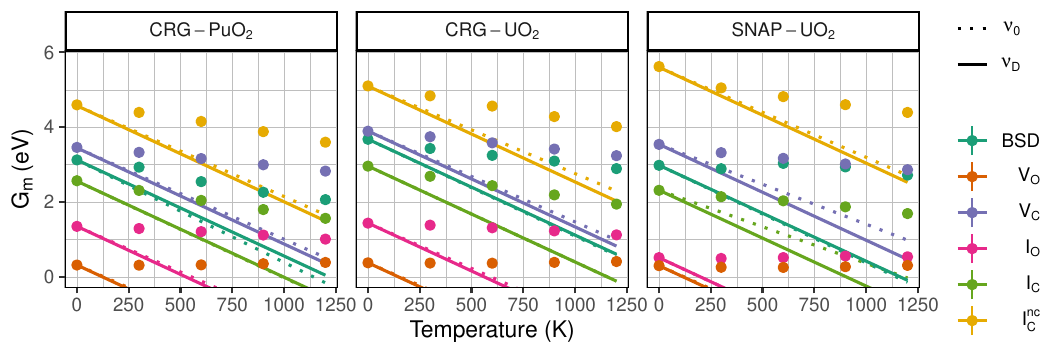}
	\caption{Migration free energy $G_\mathrm{M}$ as functions of temperature for the six defect migration mechanisms and the three interatomic potentials, using the PAFI method (symbols) and harmonic approximations (lines). Harmonic estimates are based on NEB migration barriers and either the calculated attempt frequency $\nu_0$ or the Debye frequency $\nu_\mathrm{D}$ (Table \ref{table:attfreq}).}
	\label{fig:migfreenergy}
\end{figure*}

The evolution of the free energy barrier with temperature computed using either PAFI or the harmonic approximation with the Debye or attempt frequency is shown in Fig.~\ref{fig:migfreenergy}. All of the computed free energy barriers $G_\mathrm{M}^\mathrm{PAFI}$ are higher than those determined using the two versions of the harmonic approximations. Besides, the free energy barrier $G_\mathrm{M}^\mathrm{PAFI}$ decreases at a rate lower than that predicted from the two harmonic approximations, $G_\mathrm{M}^\mathrm{harm}$ and $G_\mathrm{M}^\mathrm{Debye}$ for all potentials.
The differences between the two harmonic approximations (using the Debye frequency $\nu_\mathrm{D}$ and attempt frequency $\nu_0$) are typically quite small, despite somewhat large fluctuations in the attempt frequency between different defects. The two smallest attempt frequencies for the SNAP-UO$_2$ potential have the best agreement with the directly calculated migration barriers. This indicates that the harmonic approximation, regardless of frequency used, overestimates the contribution to the migration free energy of the attempt frequency. It also indicates that the attempt frequency is significantly lower than those calculated using the migration barrier at 0~K and is likely only applicable at temperatures less than 300~K. This can be more readily identified in Table~\ref{table:boltzmanns} which contains the migration entropy, $S_\mathrm{M}$ calculated from the direct calculation of the migration barriers and the $S_\mathrm{M}$ from each of the harmonic approximations ($\nu_\mathrm{D}$ or $\nu_0$). From these results it is apparent that for the defects with low migration barriers at 0~K the free energy of migration increases with increasing temperature (I$_\mathrm{O}$ and V$_\mathrm{O}$). 

\begin{table*}
\caption{Migration entropies $S_\mathrm{M}$ (in units of $k_\mathrm{B}$) obtained from PAFI calculations and harmonic estimates using the attempt frequency $\nu_0$. Harmonic values obtained using the Debye frequency $\nu_\mathrm{D}$ are also reported for reference.}
\label{table:boltzmanns}
\centering

\setlength{\tabcolsep}{6pt}

\begin{tabular*}{\textwidth}{@{\extracolsep{\fill}}c cc cc cc}
\toprule

& \multicolumn{2}{c}{CRG-PuO$_2$}
& \multicolumn{2}{c}{CRG-UO$_2$}
& \multicolumn{2}{c}{SNAP-UO$_2$} \\

\cmidrule(lr){2-3}
\cmidrule(lr){4-5}
\cmidrule(lr){6-7}

Defect 
& PAFI & Harmonic ($\nu_0$)
& PAFI & Harmonic ($\nu_0$)
& PAFI & Harmonic ($\nu_0$) \\

\midrule

I$_\mathrm{O}$ 
& 3.31 & 28.28 
& 3.02 & 28.11 
& -0.42 & 29.78 \\

I$_C$ 
& 9.73 & 29.94 
& 9.77 & 29.71 
& 5.85 & 22.67 \\

I$_C^{\mathrm{nc}}$ 
& 9.71 & 28.15 
& 10.54 & 27.11 
& 11.22 & 27.91 \\

BSD 
& 10.80 & 31.74 
& 7.37 & 30.02 
& 1.85 & 30.21 \\

V$_\mathrm{O}$ 
& -0.68 & 28.09 
& -0.34 & 28.11 
& -0.12 & 28.50 \\

V$_C$ 
& 6.18 & 28.30 
& 6.33 & 28.16 
& 6.40 & 24.74 \\

\cmidrule(lr){2-3}
\cmidrule(lr){4-5}
\cmidrule(lr){6-7}

Harmonic ($\nu_\mathrm{D}$)
& \multicolumn{2}{c}{29.77} 
& \multicolumn{2}{c}{29.71} 
& \multicolumn{2}{c}{29.71} \\

\bottomrule
\end{tabular*}
\end{table*}

Figure~\ref{fig:omega} shows the jump frequency ($\omega$) determined using the calculated attempt frequencies and the free energy of migration from PAFI. These are compared to the results from the harmonic approximations (utilises the free energy of migration estimated from the harmonic approximation). The jump frequencies are similar between the CRG-PuO$_2$ and CRG-UO$_2$ potentials for the oxygen defects, but vary for defects which contain the cation (BSD, I$_C$, I$_C^\mathrm{nc}$, and V$_C$). This is due to the CRG potential using the same interatomic constants for the O-O portion of the potential~\cite{cooper_many-body_2014}. The jump frequencies have significantly different slopes in the case of the SNAP-UO$_2$ potential, although the I$_C^\mathrm{nc}$ and V$_C$ defects exhibit similarity with those calculated using the CRG potential. Using the Debye or attempt frequencies again in the calculation of the jump frequency compounds the errors found when calculating the free energy of migration and the differences in jump frequency become even more obvious. These results also show that the slope is determined by the attempt frequency (or Debye frequency) whilst the intercept is attributed to the free energy of migration. 
	
\begin{figure*}[htbp]
	\label{fig:omega}
	\centering
	\includegraphics[width=\textwidth, keepaspectratio]{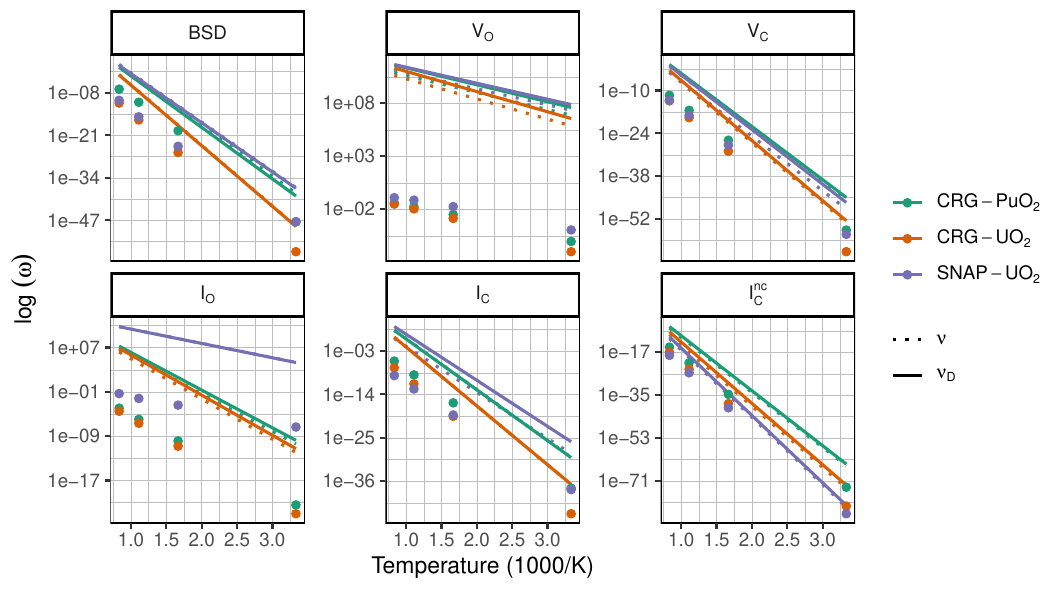}
	\caption{Jump frequency $\omega$ as a function of inverse temperature for the six defect migration mechanisms and the three interatomic potentials, calculated from PAFI migration free energies (symbols) and harmonic approximations (lines). Harmonic estimates use the migration enthalpy at 0 K combined with either the Debye frequency $\nu_\mathrm{D}$ or calculated attempt frequency $\nu_0$ (cf. Table \ref{table:attfreq}).}
	\label{fig:omega}
\end{figure*}

\section{Discussion}
\subsection{Harmonic approximation compared to fully anharmonic calculations}
The harmonic approximation using the Debye frequency is often used to estimate diffusion in materials due to the difficulty in determining the migration free energy and attempt frequency in materials with defects. New methods have been developed which probe the migration pathways derived at 0~K to determine the Gibbs free energy of migration at temperatures above 0~K. Comparing the harmonic approximations with the full anharmonic calculation of the migration free energy can indicate how well the harmonic approximation describes the phonons and diffusion in the materials, in this case UO$_2$ and PuO$_2$. 

Results found in Fig.~\ref{fig:migfreenergy} showcase the need for explicit determination of the fully anharmonic free energy of migration as a function of temperature. The results from the harmonic approximation deviate significantly from the behaviour predicted using PAFI. The differences between the harmonic approximation using the Debye frequency and attempt frequency tend to be minor except for the two oxygen defects using the SNAP-UO$_2$ potential, this is due to the much lower attempt frequency calculated for these defects. It indicates that attempt frequencies calculated at 0~K lead to an overestimation of the entropy of migration. The effects of thermal expansion and phonon-phonon interactions heavily impact the free energy of migration. Additionally, whilst the harmonic approximation would predict negative free energy of migration for the I$_\mathrm{O}$ and $V_\mathrm{O}$ defects the results from PAFI indicate the free energy of migration remains relatively stable across the temperature range, or increasing very slightly.

The jump frequencies shown in Fig.~\ref{fig:omega} also highlight the need for calculations which account for anharmonicity in order to describe the diffusion of defects in UO$_2$ and PuO$_2$. The jump frequency calculated using the two harmonic approximations are significantly different, with much higher jump frequencies, than those calculated directly using PAFI and the calculated attempt frequencies. The oxygen vacancy in particular is 10 orders of magnitude lower when directly calculated compared to the estimates found using the harmonic approximations and is similarly lower in the jump frequencies of the oxygen interstitial. The jump frequencies found for oxygen vacancies are many orders of magnitude higher than for other defect migration mechanisms at 300~K. However, other defects increase in jump frequency far more with temperature, with oxygen interstitials becoming comparable in magnitude at 1\,200~K. Additionally, at higher temperatures, the jump frequency of cation interstitials approaches that of oxygen vacancies and interstitials, being only one order of magnitude less than the jump frequency of oxygen vacancies. What is interesting to note is that the harmonic approximation is far more accurate for defects containing cations than defects containing oxygen. This highlights the issue of using the harmonic approximation, especially for oxygen defect mechanisms, without direct calculations of the jump frequencies one would assume that even at temperatures over 1\,200~K cation interstitials would still be 10 or more orders of magnitude lower than oxygen vacancies and interstitials. Direct calculations show this not to be the case.

In fact, the errors seen in the free energy of migration are compounded due to the reuse of the attempt/Debye frequency. Notably, when substituting the Debye frequency for the attempt frequency in the jump frequency calculation (when using free energy of migration calculated directly) the results are almost identical. The harmonic approximations and the directly calculated values for jump frequency have very similar gradients, however, the intercepts are significantly different between the harmonic approximation and those calculated, again due to the differences seen in the estimation of the free energy of migration. Again, it is clear that the harmonic approximation must be tested when accurate diffusion models are required at temperatures relevant to those in nuclear reactors.

\subsection{Comparison between CRG and SNAP potentials for UO$_2$}  

The comparison between the CRG and SNAP potentials highlights both agreements and important discrepancies in their description of defect migration. For most defects, the SNAP-UO$_2$ potential predicts systematically lower migration barriers than the CRG-UO$_2$ potential (see Table~\ref{table:NEB-values}), with the notable exception of the noncollinear cation interstitial (I$_C^\mathrm{nc}$), for which SNAP produces the highest barrier of all tested cases. This trend suggests that while SNAP captures the general ordering of defect mobilities, it tends to underestimate the migration enthalpy relative to CRG, except in cases where more complex cation–oxygen interactions dominate the migration mechanism.  

\begin{table*}
\caption{Migration enthalpies $\Delta H_\mathrm{M}$ (eV) obtained from NEB calculations for the six defect migration mechanisms and the three interatomic potentials. Results are compared with available density functional theory (DFT+$U$)~\cite{dorado_first-principles_2011,dorado_first-principles_2012} and experimental values from the literature~\cite{matzke_atomic_1987}.}
\label{table:NEB-values}
\centering

\setlength{\tabcolsep}{6pt}

\begin{tabular*}{\textwidth}{@{\extracolsep{\fill}}c ccc c}
\toprule

& \multicolumn{3}{c}{This work} &  \\
\cmidrule(lr){2-4}

Defect 
& CRG-PuO$_2$ & CRG-UO$_2$ & SNAP-UO$_2$ & Literature \\

\midrule

BSD 
& 3.12 & 3.68 & 2.98 & --- \\

I$_\mathrm{O}$ 
& 1.35 & 1.44 & 0.51 
& 0.93 (GGA+$U$), 0.9 (expt.) \\

I$_C$ 
& 2.56 & 2.96 & 2.32 & --- \\

I$_C^{\mathrm{nc}}$ 
& 4.57 & 5.10 & 5.60 
& 4.7 (GGA+$U$), 4.1 (LDA+$U$) \\

V$_\mathrm{O}$ 
& 0.32 & 0.38 & 0.30 
& 0.67 (GGA+$U$), 0.55 (expt.) \\

V$_C$ 
& 3.90 & 3.45 & 3.54 
& 4.8 (GGA+$U$), 3.6 (LDA+$U$), 2.4 (expt.) \\

\bottomrule
\end{tabular*}
\end{table*}
The structural features of the migration pathway also show differences between the two potentials. In particular, CRG predicts the presence of intermediate local minima for cationic defects (I$_C$, I$_C^\mathrm{nc}$, and V$_C$) associated with Frenkel pair formation, whereas SNAP yields smoother pathways without these intermediate metastable states. This absence of local minima in SNAP indicates a reduced ability to reproduce the detailed coupling between migrating cations and surrounding oxygen atoms.

Differences are also apparent in the vibrational properties. The attempt frequencies $\nu_0$ calculated with the SNAP potential span a much broader range than those of CRG, with extremely low values for some cationic defects (e.g., I$_C$ and V$_C$), contrasting with the higher and more consistent frequencies obtained with CRG. These anomalously low attempt frequencies strongly affect the estimation of migration free energies and jump frequencies when using the harmonic approximation, leading to large inconsistencies with the directly calculated PAFI results. Differences in attempt frequencies between the SNAP and CRG potential are due to the differences found in the phonon frequencies between the saddle point and initial state, with those of SNAP having smaller differences between the two states, likely attributed to the different U-U interactions in the two potentials. However, it should be noted that neither potential can be ruled more accurate in attempt frequency as this would require confirmation with computationally demanding DFT simulations. This conflicts with the SNAPs arguably more accurate reproduction of the density of states when compared to those found using experiments and DFT. In contrast, CRG produces attempt frequencies that are more uniform and closer to the expected order of magnitude derived from the Debye frequency. 

Overall, the SNAP-UO$_2$ potential provides a computationally efficient description of migration energetics and reproduces some DFT migration barriers with reasonable accuracy (e.g., I$_\mathrm{O}$ and V$_\mathrm{O}$). However, its reduced accuracy in describing cationic defect pathways, together with the large variation in attempt frequencies, indicates limitations in transferability beyond the training set. The CRG potentials, while empirical in nature, capture some qualitative features of defect migration despite the density of states differing slightly from those found using experiments and DFT. The metastable states found using the CRG potential could prove crucial in determining the potentials effectiveness at calculating properties related to defect migration. Validation of the existence of the metastable states using DFT would be a clear indicator of which potential is more effective when calculating diffusion properties. Conversely, if DFT finds no such metastable states this would be a good indicator that the SNAP potential is a better candidate for diffusion calculations. These results emphasize the need for careful validation of machine learning interatomic potentials against both DFT and experimental benchmarks before they can reliably replace traditional empirical models in defect diffusion studies.

\subsection{Comparison of defect migration in UO$_2$ and PuO$_2$}
	
The determination of attempt frequency, migration energy, migration free energy and jump frequency for both UO$_2$ and PuO$_2$ using the CRG potential allows for comparison of defect migration in UO$_2$ and PuO$_2$. Immediately noticeable is that PuO$_2$ has lower migration enthalpies, seen in Table~\ref{table:NEB-values} and Figure~\ref{fig:NEB}, which is true for all defects tested. Conversely, PuO$_2$ also has higher attempt frequencies for most defects (excluding V$_O$ which is within error of the value found for UO$_2$). This is important as the attempt frequency can compensate for the lower migration barrier energy when looking at the jump frequency and Gibbs free energy of migration, which has been observed by Messina \textit{et al.}~\cite{MESSINA2020166} following a classical Meyer-Neldel correlation~\cite{meyer1937,yelon2006}, seen in Figs.~\ref{fig:migfreenergy} and~\ref{fig:omega}. In particular, the jump frequencies of the PuO$_2$ and UO$_2$ species are very similar when comparing the directly calculated values. However, the values that were estimated using the two harmonic approximations can differ quite significantly depending on the defect under investigation. Typically the CRG-UO$_2$ potential has a lower jump frequency than PuO$_2$ but the differences are minor compared to the differences found using the harmonic approximations. The differences also tend to increase with increasing temperature. In the case of the Gibbs free energy of migration, obviously the intercept is lower for PuO$_2$, as seen in Fig.~\ref{fig:NEB}, however, the slopes calculated using the harmonic approximation are almost identical in all cases. The result is that, although the migration barrier energy is lower in PuO$_2$, this does not necessarily translate into lower diffusion coefficients or lower jump frequencies, as the increase in attempt frequencies tend to compensate for the migration barriers.

\section{Conclusions}
This research compares two distinct harmonic approximations against a direct calculation of the fully anharmonic free energy of migration in both UO$_2$ and PuO$_2$, for six different defect migration mechanisms
using the CRG-UO$_2$/CRG-PuO$_2$ potential for UO$_2$ and PuO$_2$ and a novel SNAP-UO$_2$ machine learning potential for UO$_2$. The results indicate that while the SNAP-UO$_2$ potential produces similar enthalpies of migration to the CRG-UO$_2$ potential, there are notable differences in the calculated attempt frequencies, directly calculated free energy of migration and, subsequently, the jump frequencies. Notably the SNAP potential produces lower attempt frequencies for four of the tested defect mechanisms (excluding the oxygen interstitials and bound Schottky defects), however the directly calculated jump frequencies have no concrete trend. The oxygen vacancy, bound Schottky defect, oxygen interstitial and cation vacancy have higher jump frequencies using the SNAP potential, while the two cation interstitial mechanisms exhibit for the most part lower jump frequencies when compared to the CRG potential. The differences between the two harmonic approximations used are very minor, which demonstrates that calculating the attempt frequency for use in the harmonic approximation adds little value. Conversely, the directly calculated values for the free energy of migration and jump frequency deviate from the harmonic approximation at temperatures as low as 300~K.  These differences underscore the importance of direct calculations for accurate diffusion modeling. The study also highlights the limitations of the harmonic approximation, particularly at higher temperatures. Future research should focus on validating these findings with additional DFT and experimental data to improve the understanding of defect migration in UO$_2$ and PuO$_2$. The insights gained from this study can enhance the accuracy of fuel performance codes through the utilisation of the directly calculated free energy of migration and jump frequencies which are appropriate to use when irradiation controls defect concentration and thereby contribute to the development of more reliable nuclear fuel technologies. There are numerous materials where direct calculations of the free energy of migration, jump frequency and diffusion coefficient could accelerate understanding and development, especially in sectors such as semiconductors and materials for space applications.

\section*{Acknowledgments}
This work was performed using HPC resources from GENCI TGCC Joliot-Curie on the Irene ROME partition under allocation 2025-A0180906008.
It contributes to the CEA RTA/RCOMB project and received funding from the CEA FOCUS program "Expérimentation Numérique et Jumeau Numérique" (EJN).
The authors thank Julien Tranchida for the fruitful discussions, and Petra Ospital for their contribution regarding the starting configurations.


\bibliography{main_arXiv_v1_2026-03-17Notes,References}

\appendix

\section{PAFI free energy data}
We report the full set of PAFI migration free energies for all defect migration mechanisms considered in this work. The values are computed directly from the PAFI free energy integration for each interatomic potential and temperature investigated in the main text.

Table \ref{table:PAFI-results} lists the migration free energy as a function of temperature for the different defect mechanisms and interatomic potentials studied. Results are provided for the CRG-UO$_2$, CRG-PuO$_2$, and SNAP-UO$_2$ potentials over the temperature range from 0 to 1\,200 K. These values correspond to the free energy difference between the initial state and the saddle-point configuration along the migration pathways shown in Figs.~\ref{fig:PAFI_ints} and~\ref{fig:PAFI_vacs}.

\begin{table*}[h!]
\caption{Fully anharmonic migration free energies $\Delta G_\mathrm{M}$ (eV) obtained from direct PAFI calculations for the defect migration mechanisms investigated in this work (cation interstitial I$_C$, cation interstitial with noncollinear migration I$_C^\mathrm{nc}$, oxygen interstitial I$_\mathrm{O}$, bound Schottky defect BSD, cation vacancy V$_C$, and oxygen vacancy V$_\mathrm{O}$). Values are reported as functions of temperature (in K) for the CRG-UO$_2$, CRG-PuO$_2$, and SNAP-UO$_2$ interatomic potentials.}
\label{table:PAFI-results}
\centering

\resizebox{\textwidth}{!}{
\begin{tabular*}{\textwidth}{@{\extracolsep{\fill}}c ccccc ccccc ccccc}
\toprule

& \multicolumn{5}{c}{CRG-UO$_2$}
& \multicolumn{5}{c}{CRG-PuO$_2$}
& \multicolumn{5}{c}{SNAP-UO$_2$} \\

\cmidrule(lr){2-6}
\cmidrule(lr){7-11}
\cmidrule(lr){12-16}

Defect
& 0 & 300 & 600 & 900 & 1\,200
& 0 & 300 & 600 & 900 & 1\,200
& 0 & 300 & 600 & 900 & 1\,200 \\

\midrule

I$_C$    
& 2.57 & 2.31 & 2.04 & 1.80 & 1.57
& 2.96 & 2.69 & 2.44 & 2.19 & 1.94
& 2.31 & 2.14 & 2.03 & 1.87 & 1.69 \\

I$_C^{\mathrm{nc}}$
& 4.60 & 4.39 & 4.15 & 3.88 & 3.60
& 5.10 & 4.84 & 4.56 & 4.28 & 4.01
& 5.62 & 5.05 & 4.82 & 4.60 & 4.39 \\

I$_\mathrm{O}$    
& 1.35 & 1.29 & 1.21 & 1.12 & 1.01
& 1.44 & 1.38 & 1.31 & 1.23 & 1.12
& 0.52 & 0.49 & 0.52 & 0.56 & 0.54 \\

BSD    
& 3.13 & 2.93 & 2.55 & 2.26 & 2.06
& 3.68 & 3.43 & 3.25 & 3.09 & 2.89
& 2.98 & 2.89 & 3.03 & 2.94 & 2.72 \\

V$_C$    
& 3.46 & 3.33 & 3.16 & 3.00 & 2.83
& 3.89 & 3.74 & 3.58 & 3.42 & 3.24
& 3.54 & 3.32 & 3.17 & 3.02 & 2.86 \\

V$_\mathrm{O}$    
& 0.32 & 0.31 & 0.32 & 0.35 & 0.39
& 0.38 & 0.37 & 0.37 & 0.39 & 0.41
& 0.30 & 0.26 & 0.25 & 0.27 & 0.31 \\

\bottomrule
\end{tabular*}}
\end{table*}

\end{document}